\documentclass[aps,pre,twocolumn,groupedaddress, showkeys, a4paper]{revtex4}
\usepackage{amsmath}
\usepackage{graphicx}
\usepackage{hyperref}

\begin{document}


\title{Deciphering network community structure by Surprise}

\author{Rodrigo Aldecoa}
\email[]{raldecoa@ibv.csic.es}
\author{Ignacio Mar\'in}
\email[]{imarin@ibv.csic.es}
\affiliation{Instituto de Biomedicina de Valencia.
Consejo Superior de Investigaciones Científicas (IBV-CSIC)
Calle Jaime Roig 11. Valencia, Spain}


\date{\today}

\begin{abstract}
The analysis of complex networks permeates all sciences, from biology to sociology. A fundamental, unsolved problem is how to characterize the community structure of a network. Here, using both standard and novel benchmarks, we show that maximization of a simple global parameter, which we call Surprise (S), leads to a very efficient characterization of the community structure of complex synthetic networks. Particularly, S qualitatively outperforms the most commonly used criterion to define communities, Newman and Girvan's modularity (Q). Applying S maximization to real networks often provides natural, well-supported partitions, but also sometimes counterintuitive solutions that expose the limitations of our previous knowledge. These results indicate that it is possible to define an effective global criterion for community structure and open new routes for the understanding of complex networks.
\end{abstract}


\maketitle

\section*{Introduction}
A network of interacting units is often the best abstract representation of real-life situations or experimental data. This has led to a growing interest in developing methods for network analysis in scientific fields as diverse as mathematics, physics, sociology and, most especially, biology, both to study organismic (e. g. populational, ecological) and cellular (metabolic, genomic) networks \cite{1,2,3,4,5}. A significant step to understand the properties of a network consists in determining its communities, compact clusters of densely linked, related units. However, the best way to establish the community structure of a network is still disputed. Many strategies have been used (reviewed in \cite{6}), the most popular being the maximization of Newman and Girvan's modularity (Q) \cite{7}. However, Q has the drawback of being affected by a resolution limit: its maximization fails to detect communities smaller than a threshold size that depends on the total size of the network and the pattern of connections \cite{8}. Since this finding, no other global parameters have been proposed to substitute Q. Alternative strategies (searching for local structural determinants, multilevel optimization of Q) have been suggested, but none of them has achieved general acceptance \cite{6}.

Some years ago, we suggested determining the community structure of a network by evaluating the distributions of intra- and inter-community links with a cumulative hypergeometric distribution \cite{9}. Accordingly, to find the optimal community structure of a network of symmetrically connected units (undirected graph) is equivalent to maximize the following parameter:
\begin{equation}
\label{eq1}
S=\displaystyle\sum\limits_{j=p}^{Min(M,n)} \frac{\binom{M}{j}{\binom{F-M}{n-j}}}{\binom{F}{n}}
\end{equation}
Where $F$ is the maximum possible number of links in a network (i. e. $[k^2-k]/2$, being $k$ the number of units), $n$ is the observed number of links, $M$ is the maximum possible number of intracommunity links for a given partition, and $p$ is the total number of intracommunity links actually observed in that partition. The parameter S, which stands for Surprise, indeed measures the "surprise" (improbability) of finding by chance a partition with the observed enrichment of intracommunity links in a random graph.

In this work, we show that S has features that make it the parameter of choice for global estimation of community structure. By using standard and novel benchmarks and a set of high-quality algorithms for community detection, we show that maximizing S often provides optimal characterizations of the existing communities. When this method is applied to real networks, we obtained some expected, logical solutions - some of them much better than those provided by Q maximization - but also unexpected partitions that demonstrate the limitations that the usage of inefficient tools has hitherto cast over the field.

\section*{Results}
Testing the performance of a global parameter to determine community structure requires both a set of efficient algorithms for community detection and a set of standard benchmarks, consisting in synthetic networks of known structure. In this study, six selected algorithms (see Methods) were tested in two types of benchmarks, which will be called LFR and RC throughout the text. LFR (Lancichinetti-Fortunato-Radicchi) benchmarks are characterized by providing networks in which both the degrees of the nodes and the sizes of the communities follow power laws \cite{10}. RC (Relaxed Caveman) benchmarks start with networks in which all the nodes in a community are connected. Then, this structure is relaxed by generating intercommunity links \cite{11}. We further divided LFR and RC benchmarks into "open" and "closed". Open benchmarks have been commonly used in the past (e.g. \cite{10,12,13}). In them, sets of similar networks with different proportions of intercommunity links are tested. With many intercommunity links, the networks approach randomness. In closed benchmarks, a starting community structure is progressively transformed into a second, final structure which is exactly known.

For each benchmark, we estimated S and Q with the six algorithms. The maximum values of S and Q obtained ($S_{max}$ and $Q_{max}$) provided the partitions used to compare with the known community structures. As in previous works \cite{10,14,15}, Normalized Mutual Information (NMI) was used to measure the congruence between the known and the estimated community structures. However, we also used the Variation of Information (VI) \cite{16} in a particular case.

\subsection*{Open benchmarks}
\begin{figure}[ht]
\includegraphics[scale=0.65]{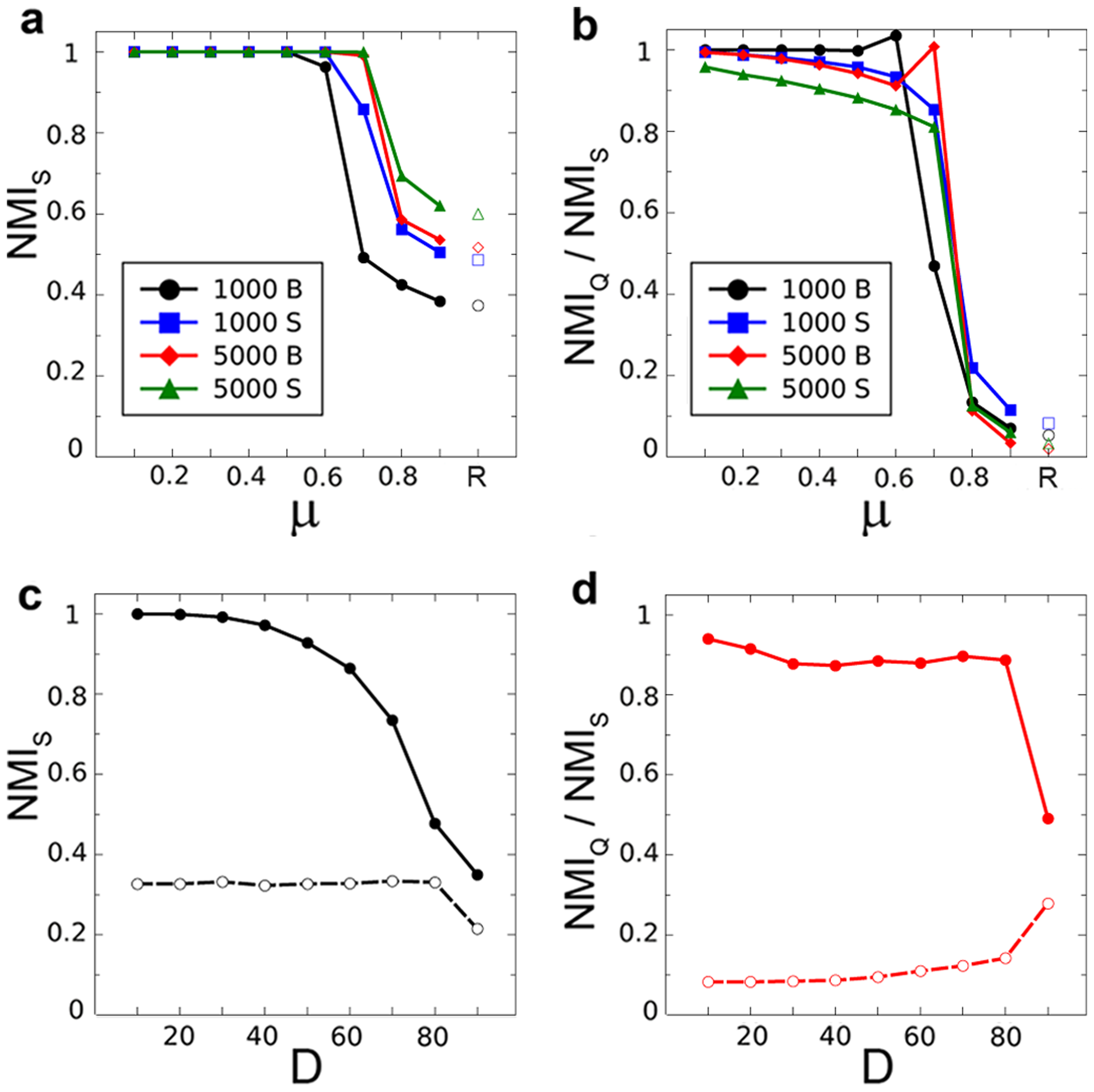}%
\caption{\label{fig:1} Results for open LFR and RC benchmarks. a) Results for the four standard LFR networks. B and S indicate big and small communities respectively and 1000 or 5000 the number of nodes. $\mu$: mixing parameter. NMI measures the congruence between the known and the deduced community structures. Each point is based on 100 different networks; standard errors of the mean are too small to be visualized. Values for 100 random (R) networks with the same number of units and degree distributions are also shown. b) Comparison of S and Q maximizations in LFR benchmarks. The NMI$_Q$/NMI$_S$ ratios, which are almost always below 1, are shown. c) Results for the RC benchmark. The parameter Degradation (D) indicates the percentage of both deleted and shuffled links. Each black dot is based on 100 networks, again standard errors are so small that cannot be visualized at this scale. For each value of D, results for 100 random networks with the same number of links are also shown (open circles). d) Relative quality of the partitions generated by maximizing S and Q in RC benchmarks. As in panel b, NMI$_Q$/NMI$_S$ ratios are shown. White dots: results for random networks with different D values.}
\end{figure}
\begin{figure}[ht]
\includegraphics[scale=0.95]{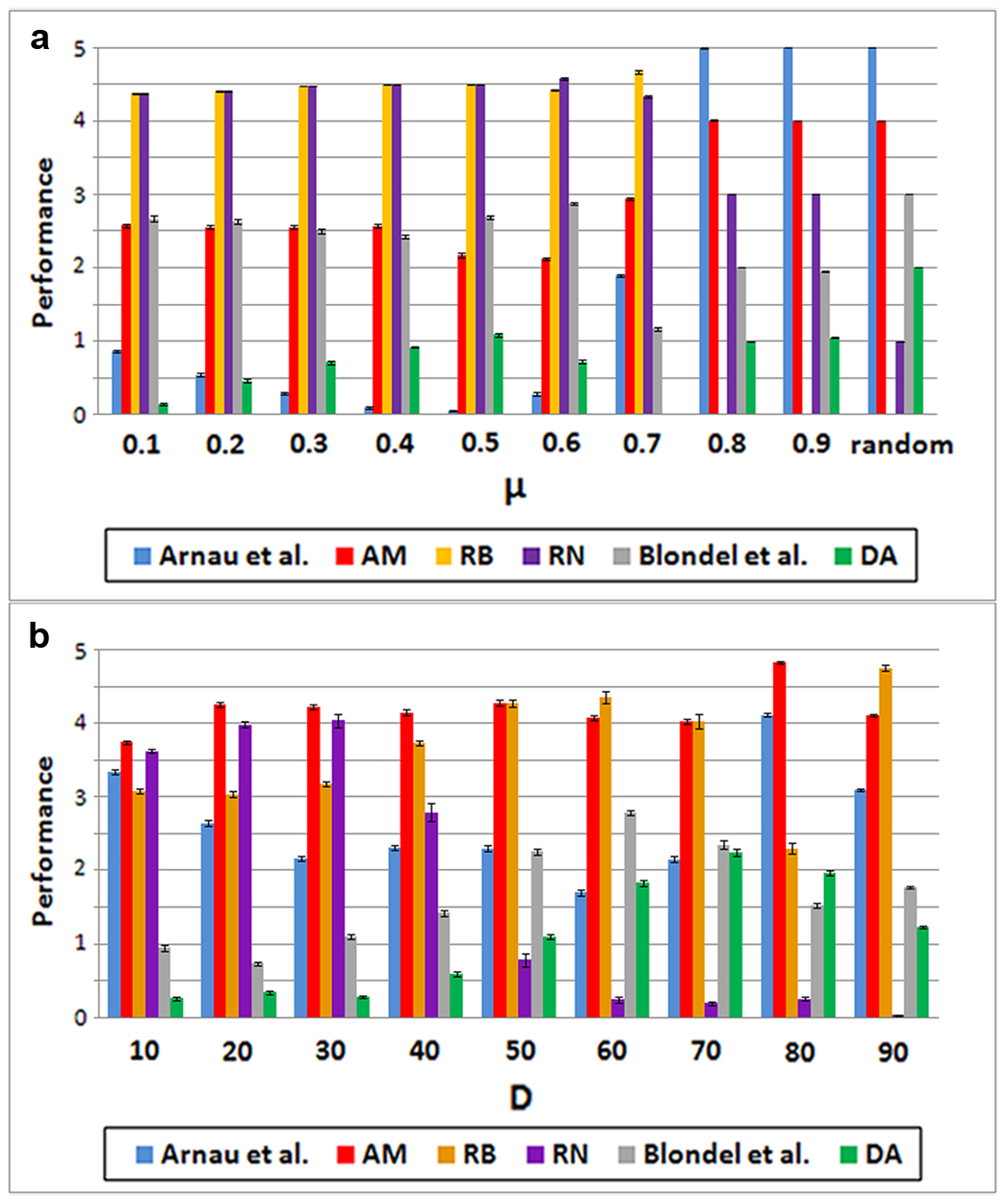}%
\caption{\label{fig:2} Average performance of the algorithms in the open LFR and RC benchmarks. The algorithms used were described by Arnau et al. \cite{9}, Aldecoa and Marín (AM) \cite{13}, Rosvall and Bergstrom (RB) \cite{23}, Ronhovde and Nussinov (RN) \cite{24}, Blondel et al. \cite{25} and Duch and Arenas (DA) \cite{26}. a) Typical example of the results obtained in LFR benchmarks, here with 5000 units and big communities (see Figure S1 for all of them). After ordering the algorithms from best to worst performance, their ranks were added for the 100 different networks. Performance was defined as $P = 6 - average rank$. Therefore, the maximum value $P = 5$ means that an algorithm was the best in all networks tested, while $P = 0$ means that it was always the worst. As it can be observed, none of the algorithms achieved optimal results in all cases. b) Results obtained in the RC benchmark with different Degradation (D) values. Performance evaluated as in panel a).}
\end{figure}

Figures \ref{fig:1}a and \ref{fig:1}b summarize the results obtained for four standard open LFR benchmarks that differ in number of units and community sizes \cite{10} (see Methods). Figure \ref{fig:1}a indicates that selecting the solution with a maximum S value leads to a perfect characterization of the network structure (NMI$_S = 1$) even when that structure is blurred by a large number of inter-community links, generated by increasing the mixing parameter $\mu$ up to 0.5-0.7 (see Methods for $\mu$ definition). If $\mu$ is further increased, the original partition is not chosen by any algorithm (NMI$_S < 1$). This suggests that the original community structure is not present anymore, which is in good agreement with the fact that $S_{max} \gg S_{orig}$, where $S_{orig}$ is the S value obtained assuming that the original community structure is still present (Table S1). S maximization qualitatively improves over Q maximization (Figure 1b and Table S1): NMI$_S$ $>$ NMI$_Q$ in $2827/3600 = 78.5\%$ of the cases, NMI$_Q$ $>$ NMI$_S$ in just 4.1\% of them and the rest are ties. Interestingly, NMI$_Q$ $\ll$ NMI$_S$ in quasi-random and random networks (Figure 1b), suggesting that maximizing Q overimposes spurious community structures in those cases. It is significant that S maximization provided better average NMI scores than those obtained by any single algorithm in these same benchmarks \cite{15}. Different algorithms provided the top S scores, depending on the benchmark and $\mu$ value examined (Figure \ref{fig:2}a and Figure S1).

The discovery of the resolution limit of Q showed that heterogeneous community sizes may greatly affect the ability of global parameters to detect structure \cite{8}. However, by construction, community sizes in the standard LFR benchmarks are very similar. Pielou's evenness indexes (PI) \cite{17} ranged from 0.96 to 0.98 in the four benchmarks used above, close to the maximum value of the index (PI = 1 for communities of identical size). Considering that it was critical to test S in more extreme situations, we built the RC benchmarks, which have PIs as low as 0.70 (as shown in Figure S2). Figures \ref{fig:1}c and \ref{fig:1}d summarize the results for open RC benchmarks, with progressive Degradation (D; see Methods) of the original structure. That structure is efficiently detected by S maximization, with a slow decrease in performance when D increases (Figure 1c; see also Table S2, Figure S2). Again, S maximization clearly improves over Q maximization in these benchmarks (Figure \ref{fig:1}d; NMI$_S$ $>$ NMI$_Q$ in $848/900 = 94.2\%$ of the cases, while NMI$_Q >$ NMI$_S$ in just 3.3\% of the cases). As occurred for the LFR benchmarks, none of the algorithms obtained the best results in all networks (Figure \ref{fig:2}b).

\subsection*{Closed benchmarks}
\begin{figure}[p]
\includegraphics[scale=0.85]{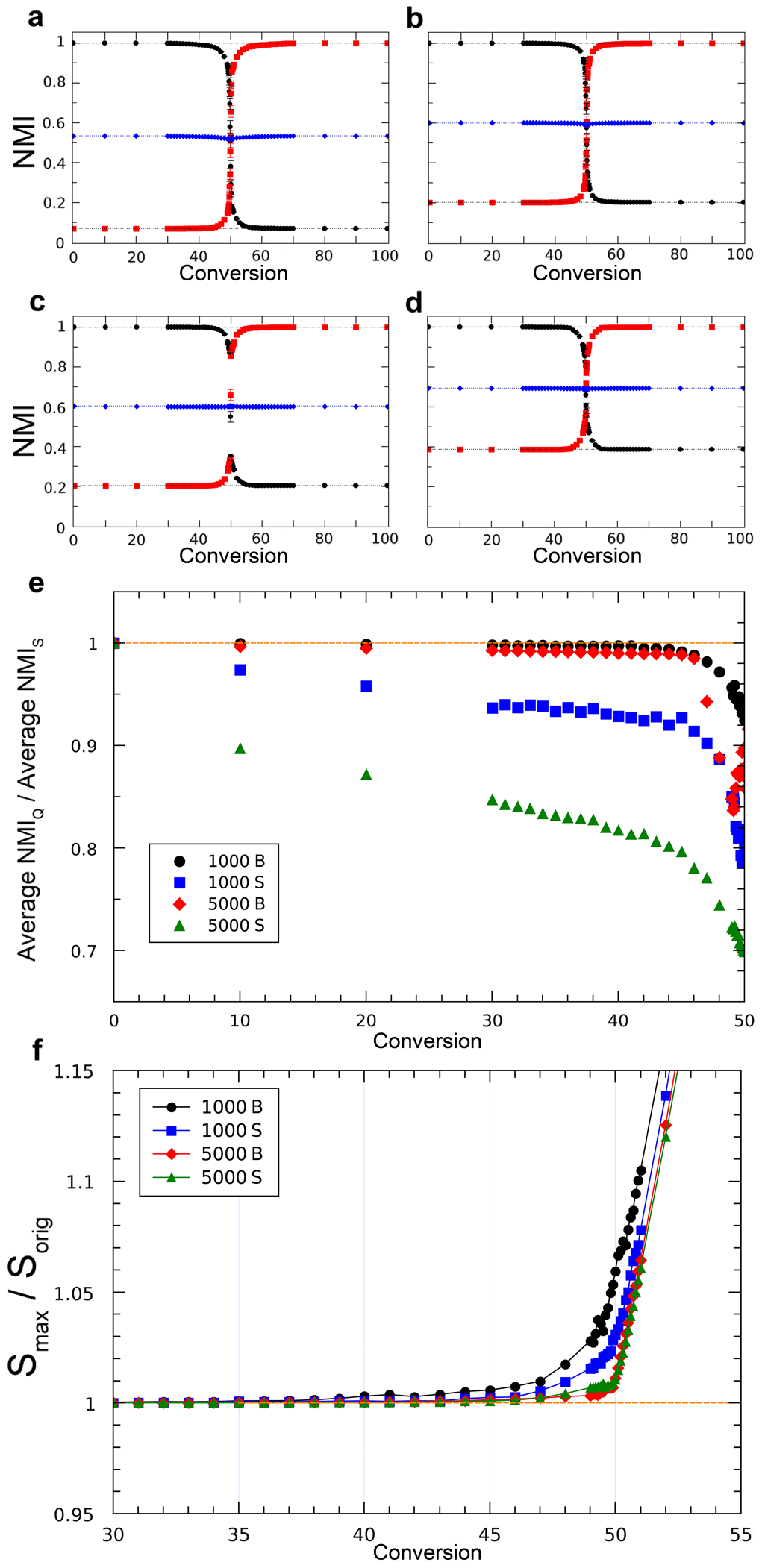}%
\caption{\label{fig:3} Results for closed LFR benchmarks. a) LFR benchmark with 1000 units and big communities. For each Conversion (C) value, NMIs comparing the $S_{max}$ partition with the initial (black dots) or final (red squares) community structures were obtained. The symmetrical results led to NMI averages (blue diamonds) that, with great precision, fell in a straight line of value $(1+$NMI$_{IF})/2$. Dots are based on 100 independent analyses. b–d) LFR benchmarks with, respectively, 1000 units, small communities (b), 5000 units, big communities (c) and 5000 units, small communities (d). Results are very similar to those in panel a). e) Average NMI values for partitions obtained maximizing Q are worse than those obtained maximizing S, especially as we move towards $C = 50$, in which the real community structure is more difficult to establish. This effect is exacerbated by large number of units and small community sizes, due to the resolution limit of Q. Results for $C > 50$ are symmetrical to the ones shown here. See also Figure S3. f) $S_{max}/S_{orig}$ ratio $\geq1$, i. e. either the original structure or a different one with higher S is found. These results are compatible with the algorithms used being able to detect the true structure present with great accuracy.}
\end{figure}

\begin{figure}[p]
\includegraphics[scale=0.85]{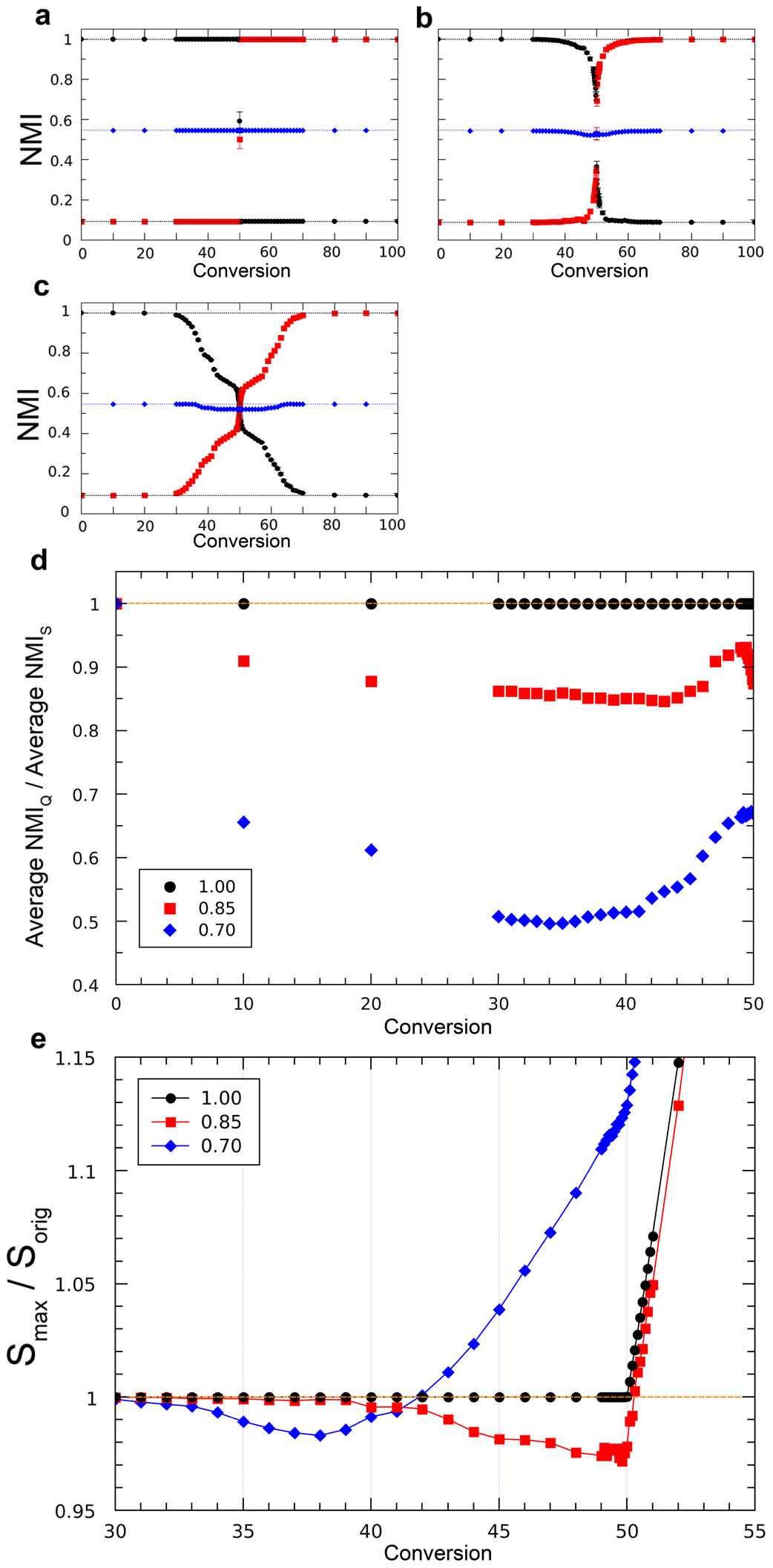}%
\caption{\label{fig:4} Results for closed RC benchmarks. Three networks with different heterogeneity in community sizes (Pielou's indexes equal to 0.70, 0.85 and 1.00 respectively) were used as examples. a) $PI = 1$; b) $PI = 0.85$; c) $PI = 0.70$. Results similar to those in Figure 2, except that the figures are not so perfectly symmetrical in the most heterogeneous networks (panels b and c; blue diamonds slightly deviate from the straight line). d) Average NMI values are much worse when Q is used, provided that community sizes are heterogeneous. See also Figure S4. e) $S_{max}/S_{orig} < 1$ with heterogeneous community sizes. The algorithms used did not detect in those cases the maximum possible S, which still may correspond to the initial structure. This may contribute to the departures from symmetry shown in panel a). The fact that $S_{max}/S_{orig} \gg 1$ with $C < 0.50$ and $PI = 0.70$ (blue diamonds) implies that the algorithms are detecting structures different from the initial one.}	
\end{figure}

The results just shown indicate that using $S_{max}$ to detect community structure has obvious advantages over maximizing Q. However, they do not allow to evaluate how optimal is that criterion, given that the potential maximum NMIs are unknown. To solve this limitation, we generated closed LFR and RC benchmarks, in which we had an \textit{a priori} expectation of the maximum NMI values. Results are shown in Figures \ref{fig:3} (LFR) and \ref{fig:4} (RC). In all cases in which $S_{max}$ was used, an almost perfectly symmetrical dynamics was observed. In the process of converting the original structure into the final one (by increasing the Conversion parameter; see Methods), NMI losses for the first structure are compensated by increases for the second. The average of both NMIs is thus approximately constant, and it has a value identical or very close to ($1+$NMI$_{IF})/2$, where NMI$_{IF}$ is obtained comparing the initial and final structures (Figures \ref{fig:3}a-d; Figures \ref{fig:4}a-c; Figures S3, S4). This is exactly the result expected for an optimal parameter (see theoretical details in Methods). On the contrary, maximizing Q shows a poor performance except when community sizes are very similar/identical (Figures \ref{fig:3}e, \ref{fig:4}d; Figures S3, S4). The same results were obtained using a second measure of congruence, Variation of Information (VI) (Figures S5, S6). Finally, in the LFR benchmarks, $S_{max}$ was always identical or higher than $S_{orig}$ (Figure \ref{fig:3}f). However, this does not happen for the RC benchmarks (Figure \ref{fig:4}e). Therefore, these algorithms sometimes fail to obtain the highest possible S values. This fact may explain the slight departures from NMI symmetry observed in some RC benchmarks (blue diamonds in Figures \ref{fig:4}b, \ref{fig:4}c).

\subsection*{Real networks}
\begin{figure}[ht]
\includegraphics[scale=0.75]{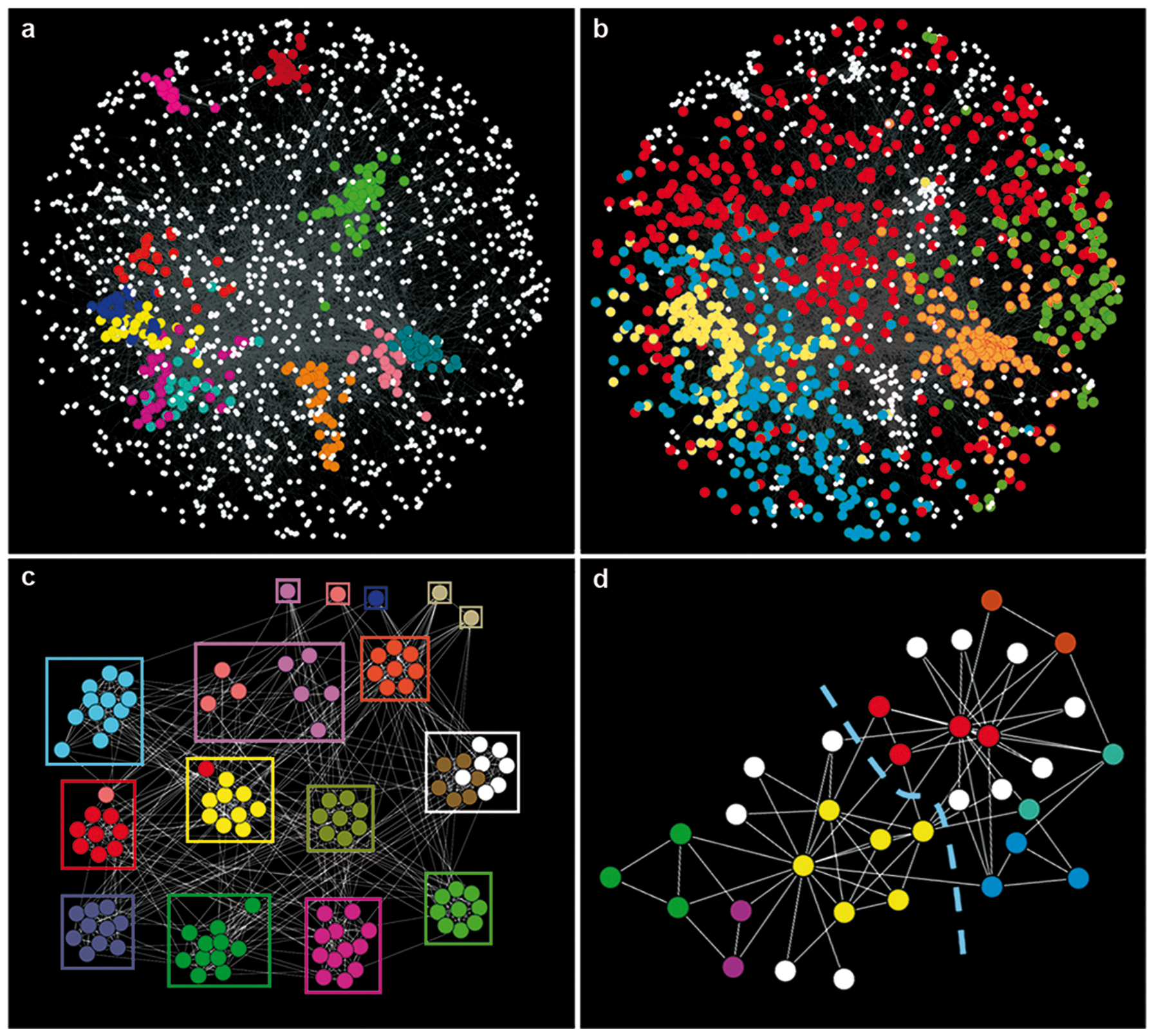}%
\caption{\label{fig:5} Community structure of the CYC2008 network (a, b), College football network (c) and Zachary's karate club network (d), according to S maximization (panels a, c, d) or Q maximization (panel b). In panel c, the known community structure is shown (squares). The broken lines in panel d divide the network into the two communities assumed to exist. That division of the network is not supported at all by $S_max$ analyses. While $S_{(2 communities)}$ = 13.61, the optimal division found has $S_{(19 communities)}$ = 25.69. Twelve of these optimal communities are singletons (white dots).}
\end{figure}

Figure 5 summarizes the Smax results for three real networks. The first example is based on the CYC2008 database, which compiles 1604 proteins that belong to 324 protein complexes \cite{18}. The general agreement between communities detected using $S_{max}$ and \textit{a priori} defined protein complexes is almost perfect, NMI$_S$ = 0.91. On Figure \ref{fig:5}a, the 11 communities of size $>$20, out of the 313 detected, are detailed to show how fine-grained is the classification obtained. On the contrary, optimizing Q provides a very coarse classification into just 24 communities with NMI$_Q$ = 0.57. The largest five communities alone almost cover the whole network (Figure \ref{fig:5}b). These results indicate how excellent is S performance when there are many small, abundant communities, a typical situation in which Q, affected by its resolution limit, radically fails. Figure \ref{fig:5}c shows, as a positive control, the results for a classical benchmark of well-known structure, the \textit{College football network} \cite{12}. The agreement with the expected communities is again very high (NMI$_S$ = 0.93). Finally, Figure \ref{fig:5}d shows the results for another well-known example, the \textit{Zachary's Karate club} network \cite{12,19}. This social network supposedly contains two communities. However, S analyses surprisingly unearthed 19 communities, 12 of them singletons (Figure \ref{fig:5}d).

\section*{Discussion}
In this study, we have shown the potential of maximizing the global parameter Surprise (S) to determine the community structure present in complex networks. The results indicate that it has a qualitative better performance than the hitherto most commonly used global measure, Newman and Girvan's modularity (Q). The advantage of S over Q is maybe not that surprising, considering the different theoretical foundations of both measures. Newman and Girvan's Q is based on a simple definition of community, as a region of the network with an unexpectedly high density of links. However, the number of units within each community does not influence the value of Q \cite{7}. On the contrary, S evaluates both the number of links and of units in each community (see \ref{eq1}). Therefore, S implicitly assumes a more complex definition of community: a precise number of units for which it is found a density of links which is statistically unexpected given the features of the network. In this context of comparison of both measures, it is also very significant that, while some of the algorithms used in this work were the best among those specifically designed to maximize Q, none was devised to maximize S. Therefore, our results actually underestimate the power of S maximization for community detection. A direct example of that underestimation is shown in Figure \ref{fig:4}e: the maximum values of S were, in some cases, not found. The few exceptions found in which NMI$_Q$ $>$ NMI$_S$ (3-4\% of all the cases examined in the open benchmarks) could be also explained by an incomplete success in determining $S_{max}$ with these algorithms.

The commonly used open benchmarks are useful for general evaluations of the performance of different algorithms, but they do not allow to establish how optimal are the results obtained. For that, we have devised novel closed benchmarks in which an initial known community structure is progressively transformed into a second, also known, community structure. Provided that both community structures are identical, it can be demonstrated that, at any point of the transformation from one to the other, the average of the NMIs of the solution found respect to the initial and final structures should approximate a constant value ($[1+$NMI$_{IF}]/2$), if that solution is optimal (see Methods). This feature allows establishing the intrinsic quality of the partitions obtained, with S maximization often providing optimal results. We conclude that S maximization establishes the community structure of complex networks with a high accuracy. Two promising lines of research are clear. First, generating novel, specific algorithms for S maximization, which may improve over the existing ones. Second, building a standard set of closed benchmarks to test any new algorithms for community detection. Our LFR and RC closed benchmarks may be a good starting point for that standard set.

When S maximization was applied to real networks, the results obtained are of two types. On one hand, for the CYC2008 and College football networks, the expectation was to find a clear community structure which should faithfully correspond to either the complexes to which the proteins examined are part (CYC2008 network) or to the conferences to which the teams belong (\textit{College football network}), given that intracomplex or intraconference links are abundant (e. g. Figure \ref{fig:5}c). These are exactly the results found using $S_{max}$. On the other hand, the structure of the \textit{Zachary's karate} network is far from obvious (Figure \ref{fig:5}d). Therefore, finding that, according to $S_{max}$, the network contains some small groups plus many singletons is, at least \textit{a posteriori}, not so unexpected. A natural question is then why the scientific community has been so keen of exploring this particular network, often to establish whether an algorithm was able or not to detect the putative two communities (e. g. refs. \cite{7,12,19,20} among many others). This may reflect a psychological bias, to which the use of underperforming methods for community detection may have certainly contributed. It shows to which extent human prejudices may taint evaluations in this type of ill-defined problems.

\section*{Methods}
\subsection*{Algorithms used to maximize S and Q}
Six of the best available algorithms, selected either by their exceptional performance in artificial benchmarks or their success in previous analyses of real and simulated networks \cite{9,13,14,15,21,22}, were used. They were the following: 1) UVCluster algorithm \cite{9,13}: It performs iterative hierarchical clustering, generating dendrograms. The best values of S and Q were obtained scanning these dendrograms from root to leaves. 2) SCluster algorithm \cite{13}: also performs iterative hierarchical clustering, but using an alternative strategy which is faster and sometimes more accurate than the one implemented in UVCluster. 3) Dynamic algorithm by Rosvall and Bergstrom \cite{23}: an algorithm based on expressing the characterization of communities as an information compression problem. 4) Potts model multiresolution algorithm \cite{24}: works by minimizing the Hamiltonian of a Potts spin model at different resolution scales, i. e. searching for communities of different sizes. 5) Fast modularity optimization \cite{25}: devised to maximize Q. It provides multiple solutions from which values for S and Q can be obtained, and the maximum ones were used in our analyses. 6) Extremal optimization algorithm \cite{26}: A divisive algorithm also developed to maximize Q. Analyses were always performed with the default program settings.

\subsection*{Features of the benchmarks}
First, the recently developed LFR benchmarks, specifically devised for testing alternative community detection strategies \cite{10}, were used. In particular, we chose four standard LFR benchmarks already explored by other authors \cite{15}. The networks analyzed had either 1000 or 5000 units and were built according to two alternative ranges of community sizes (Big (B): 20-100 units/community; Small (S): 10-50 units/community). For each of the four conditions (1000 B, 1000 S, 5000 B, 5000 S), 100 different networks were generated for each value of a mixing parameter $\mu$, which varied from 0.1 to 0.9 \cite{15}. $\mu$ is the average percentage of links that connect a unit to those in other communities. Logically, increasing $\mu$ weakens the network community structure. When $\mu$ = 0.9, the networks are quasi-random (see below).

Once found that these LFR benchmarks generated networks with communities of very similar sizes, we decided to implement RC benchmarks in which these sizes were more variable. All networks in these benchmarks had 512 units divided into 16 communities. One hundred networks with random community sizes, determined using a broken-stick model \cite{27}, were generated. This model provides highly heterogeneous community sizes. Progressive weakening of the community structure of the RC networks, similar to the effect of increasing $\mu$ in the LFR networks, was obtained as follows. Initially, all units of each community in the network were fully connected. Then, that obvious structure was progressively blurred, by first randomly removing a certain percentage of edges and then randomly shuffling the same percentage of links among the units. That common percentage, we have called Degradation (D). Thus, D = 10\% means that, first, 10\% of the links present were eliminated and then 10\% of the remaining edges were randomly shuffled among units. Shuffling involved first the random removal of an edge of the graph and then the addition of a new edge between two randomly chosen nodes.

In the LFR and RC benchmarks just described it was possible to compare networks having obvious community structures (generated with low $\mu$ or D parameters) with others that were increasingly random. This type of benchmarks, we have called open. We also generated closed LFR and RC benchmarks. In them, links were shifted in a directed way, in order to convert the original community structure of a network into a second, also predefined, structure. In this way, it is possible to monitor when the original structure is substituted by the final one according to the solutions provided by $S_{max}$ or $Q_{max}$. In the LFR and RC closed benchmarks, the starting networks were the same described in the previous paragraphs, with $\mu$ = 0.1 (LFR) or D = 0 (RC) respectively, and the final networks were obtained by randomly relabeling the nodes. Therefore, the initial and final networks had identical community structures but the nodes within each community were different. Conversion (C) is defined as the percentage of links exclusively present in the initial network that are substituted by links only present in the final one (i. e. C = 0: initial structure present; C = 100: final structure present).

\subsection*{NMI symmetry as a measure of performance in closed benchmarks}
In our closed benchmarks, a peculiar symmetrical behavior of NMI values respect to the initial and final partitions is expected. Imagine that a putative optimal partition is estimated according to a given criterion. Let us now consider the following triangle inequality:

\begin{equation}
\label{eq2}
\displaystyle\frac{NMI_{IE}+NMI_{EF}}{2} \leq \frac{1+NMI_{IF}}{2}
\end{equation}

where NMI$_{IE}$ is the normalized mutual information calculated for the initial structure (I) and the estimated partition (E), NMI$_{EF}$ is the normalized mutual information for the final structure (F) versus the estimated partition and NMI$_{IF}$ is the normalized mutual information for the comparison between the initial and final structures. Inequality \ref{eq2} holds true if the structures of I, F and E are identical (i. e. both the number and sizes of the communities are the same, but not necessarily are the same the nodes within each community). This follows from the fact that

\begin{equation}
\label{eq3}
\displaystyle{1+NMI_{XY}} \leq \frac{VI_{XY}}{H(X)+H(Y)}
\end{equation}

Where $VI_{XY}$ is the Variation of Information for both partitions \cite{16} and H(X) and H(Y) are the entropies of the X and Y partitions, respectively. Given that VI is a metric \cite{16}, it satisfies the triangle inequality

\begin{equation}
\label{eq4}
\displaystyle VI_{AB} + VI_{BC} \geq VI_{AC}
\end{equation}

If, as indicated, the structures of all partitions are identical, then all their entropies are also identical. In that case, the following inequality can be deduced from formulae \ref{eq3} and \ref{eq4}

\begin{equation}
\label{eq5}
\displaystyle (1-NMI_{AB}) + (1-NMI_{BC}) \geq (1-NMI_{AC})
\end{equation}

From this inequality, and substituting A, B and C with I, E and F, respectively, formula \ref{eq2} can be deduced. Formula \ref{eq2} therefore means that, provided that I, E and F have the same structure, the average of NMI$_{IE}$ and NMI$_{EF}$ may acquire a maximum value [$(1+$NMI$_{IF})/2$]. Inequality \ref{eq2} will also hold approximately true if the entropies of I, E and F are very similar (i. e. many identical communities). In our closed benchmarks the I and F structures are identical, and we progressively convert one into the other. It is thus expected that the optimal partition along this conversion is similar in structure to both I and F. Hence, deviations from the expected average value $(1+$NMI$_{IF})/2$ are a cause of concern, as they probably mean that the optimal partition has not been found. On the other hand, finding values equal to $(1+$NMI$_{IF})/2$ is a strong indication that the optimal partition has indeed been found.

It is worth noting that, although NMI has been commonly used in this field \cite{10,14,15}, using VI instead has clear advantages to analyze closed benchmarks: Formula \ref{eq4} can be used instead of Formula \ref{eq2}, avoiding considering entropies at all. This is why we evaluated the closed benchmark results both using NMI and VI (see above).

\subsection*{Real networks}
Two of the three networks explored, known as \textit{College football} and \textit{Zachary's karate} networks, have been frequently used in the past in the context of community detection [e. g. refs. \cite{7,12,19,20,28}. The third network derived from the CYC2008 protein complexes database \cite{18}. This database contains information for 408 protein complexes of the yeast Saccharomyces cerevisiae. The protein complex data were converted into 324 non-overlapping complexes by assigning each protein present in multiple complexes to the largest one. This was made to allow for NMI calculations. Once each protein (unit) was assigned to a non-overlapping cluster (community), we downloaded from the BioGRID database [29] the protein-protein interactions (edges) characterized so far for all these proteins. The final graph contained 1604 nodes and 14171 edges.

\bibliography{Surprise}

\begin{thebibliography}{28}
\expandafter\ifx\csname natexlab\endcsname\relax\def\natexlab#1{#1}\fi
\expandafter\ifx\csname bibnamefont\endcsname\relax
  \def\bibnamefont#1{#1}\fi
\expandafter\ifx\csname bibfnamefont\endcsname\relax
  \def\bibfnamefont#1{#1}\fi
\expandafter\ifx\csname citenamefont\endcsname\relax
  \def\citenamefont#1{#1}\fi
\expandafter\ifx\csname url\endcsname\relax
  \def\url#1{\texttt{#1}}\fi
\expandafter\ifx\csname urlprefix\endcsname\relax\def\urlprefix{URL }\fi
\providecommand{\bibinfo}[2]{#2}
\providecommand{\eprint}[2][]{\url{#2}}

\bibitem[{\citenamefont{Barab{\'a}si and Oltvai}(2004)}]{1}
\bibinfo{author}{\bibfnamefont{A.}~\bibnamefont{Barab{\'a}si}}
  \bibnamefont{and} \bibinfo{author}{\bibfnamefont{Z.}~\bibnamefont{Oltvai}},
  \bibinfo{journal}{Nature Reviews Genetics} \textbf{\bibinfo{volume}{5}},
  \bibinfo{pages}{101} (\bibinfo{year}{2004}).

\bibitem[{\citenamefont{Wasserman and Faust}(1994)}]{2}
\bibinfo{author}{\bibfnamefont{S.}~\bibnamefont{Wasserman}} \bibnamefont{and}
  \bibinfo{author}{\bibfnamefont{K.}~\bibnamefont{Faust}},
  \emph{\bibinfo{title}{Social network analysis: Methods and applications}},
  vol.~\bibinfo{volume}{8} (\bibinfo{publisher}{Cambridge university press},
  \bibinfo{year}{1994}).

\bibitem[{\citenamefont{Strogatz}(2001)}]{3}
\bibinfo{author}{\bibfnamefont{S.}~\bibnamefont{Strogatz}},
  \bibinfo{journal}{Nature} \textbf{\bibinfo{volume}{410}},
  \bibinfo{pages}{268} (\bibinfo{year}{2001}).

\bibitem[{\citenamefont{Costa et~al.}(2007)\citenamefont{Costa, Rodrigues,
  Travieso, and Boas}}]{4}
\bibinfo{author}{\bibfnamefont{L.}~\bibnamefont{Costa}},
  \bibinfo{author}{\bibfnamefont{F.}~\bibnamefont{Rodrigues}},
  \bibinfo{author}{\bibfnamefont{G.}~\bibnamefont{Travieso}}, \bibnamefont{and}
  \bibinfo{author}{\bibfnamefont{P.}~\bibnamefont{Boas}},
  \bibinfo{journal}{Advances in Physics} \textbf{\bibinfo{volume}{56}},
  \bibinfo{pages}{167} (\bibinfo{year}{2007}).

\bibitem[{\citenamefont{Newman}(2010)}]{5}
\bibinfo{author}{\bibfnamefont{M.}~\bibnamefont{Newman}},
  \emph{\bibinfo{title}{Networks: an introduction}} (\bibinfo{publisher}{Oxford
  University Press, Inc.}, \bibinfo{year}{2010}).

\bibitem[{\citenamefont{Fortunato}(2010)}]{6}
\bibinfo{author}{\bibfnamefont{S.}~\bibnamefont{Fortunato}},
  \bibinfo{journal}{Physics Reports} \textbf{\bibinfo{volume}{486}},
  \bibinfo{pages}{75} (\bibinfo{year}{2010}).

\bibitem[{\citenamefont{Newman and Girvan}(2004)}]{7}
\bibinfo{author}{\bibfnamefont{M.}~\bibnamefont{Newman}} \bibnamefont{and}
  \bibinfo{author}{\bibfnamefont{M.}~\bibnamefont{Girvan}},
  \bibinfo{journal}{Physical review E} \textbf{\bibinfo{volume}{69}},
  \bibinfo{pages}{026113} (\bibinfo{year}{2004}).

\bibitem[{\citenamefont{Fortunato and Barthelemy}(2007)}]{8}
\bibinfo{author}{\bibfnamefont{S.}~\bibnamefont{Fortunato}} \bibnamefont{and}
  \bibinfo{author}{\bibfnamefont{M.}~\bibnamefont{Barthelemy}},
  \bibinfo{journal}{Proceedings of the National Academy of Sciences}
  \textbf{\bibinfo{volume}{104}}, \bibinfo{pages}{36} (\bibinfo{year}{2007}).

\bibitem[{\citenamefont{Arnau et~al.}(2005)\citenamefont{Arnau, Mars, and
  Mar{\'\i}n}}]{9}
\bibinfo{author}{\bibfnamefont{V.}~\bibnamefont{Arnau}},
  \bibinfo{author}{\bibfnamefont{S.}~\bibnamefont{Mars}}, \bibnamefont{and}
  \bibinfo{author}{\bibfnamefont{I.}~\bibnamefont{Mar{\'\i}n}},
  \bibinfo{journal}{Bioinformatics} \textbf{\bibinfo{volume}{21}},
  \bibinfo{pages}{364} (\bibinfo{year}{2005}).

\bibitem[{\citenamefont{Lancichinetti et~al.}(2008)\citenamefont{Lancichinetti,
  Fortunato, and Radicchi}}]{10}
\bibinfo{author}{\bibfnamefont{A.}~\bibnamefont{Lancichinetti}},
  \bibinfo{author}{\bibfnamefont{S.}~\bibnamefont{Fortunato}},
  \bibnamefont{and} \bibinfo{author}{\bibfnamefont{F.}~\bibnamefont{Radicchi}},
  \bibinfo{journal}{Physical Review E} \textbf{\bibinfo{volume}{78}},
  \bibinfo{pages}{046110} (\bibinfo{year}{2008}).

\bibitem[{\citenamefont{Watts}(2003)}]{11}
\bibinfo{author}{\bibfnamefont{D.}~\bibnamefont{Watts}},
  \emph{\bibinfo{title}{Small worlds: the dynamics of networks between order
  and randomness}} (\bibinfo{publisher}{Princeton university press},
  \bibinfo{year}{2003}).

\bibitem[{\citenamefont{Girvan and Newman}(2002)}]{12}
\bibinfo{author}{\bibfnamefont{M.}~\bibnamefont{Girvan}} \bibnamefont{and}
  \bibinfo{author}{\bibfnamefont{M.}~\bibnamefont{Newman}},
  \bibinfo{journal}{Proceedings of the National Academy of Sciences}
  \textbf{\bibinfo{volume}{99}}, \bibinfo{pages}{7821} (\bibinfo{year}{2002}).

\bibitem[{\citenamefont{Aldecoa and Mar{\'\i}n}(2010)}]{13}
\bibinfo{author}{\bibfnamefont{R.}~\bibnamefont{Aldecoa}} \bibnamefont{and}
  \bibinfo{author}{\bibfnamefont{I.}~\bibnamefont{Mar{\'\i}n}},
  \bibinfo{journal}{PloS one} \textbf{\bibinfo{volume}{5}},
  \bibinfo{pages}{e11585} (\bibinfo{year}{2010}).

\bibitem[{\citenamefont{Danon et~al.}(2005)\citenamefont{Danon,
  D{\'\i}az-Guilera, Duch, and Arenas}}]{14}
\bibinfo{author}{\bibfnamefont{L.}~\bibnamefont{Danon}},
  \bibinfo{author}{\bibfnamefont{A.}~\bibnamefont{D{\'\i}az-Guilera}},
  \bibinfo{author}{\bibfnamefont{J.}~\bibnamefont{Duch}}, \bibnamefont{and}
  \bibinfo{author}{\bibfnamefont{A.}~\bibnamefont{Arenas}},
  \bibinfo{journal}{Journal of Statistical Mechanics: Theory and Experiment}
  \textbf{\bibinfo{volume}{P09008}} (\bibinfo{year}{2005}).

\bibitem[{\citenamefont{Lancichinetti and Fortunato}(2009)}]{15}
\bibinfo{author}{\bibfnamefont{A.}~\bibnamefont{Lancichinetti}}
  \bibnamefont{and}
  \bibinfo{author}{\bibfnamefont{S.}~\bibnamefont{Fortunato}},
  \bibinfo{journal}{Physical Review E} \textbf{\bibinfo{volume}{80}},
  \bibinfo{pages}{056117} (\bibinfo{year}{2009}).

\bibitem[{\citenamefont{Meil{\u{a}}}(2007)}]{16}
\bibinfo{author}{\bibfnamefont{M.}~\bibnamefont{Meil{\u{a}}}},
  \bibinfo{journal}{Journal of Multivariate Analysis}
  \textbf{\bibinfo{volume}{98}}, \bibinfo{pages}{873} (\bibinfo{year}{2007}).

\bibitem[{\citenamefont{Rosvall and Bergstrom}(2008)}]{23}
\bibinfo{author}{\bibfnamefont{M.}~\bibnamefont{Rosvall}} \bibnamefont{and}
  \bibinfo{author}{\bibfnamefont{C.}~\bibnamefont{Bergstrom}},
  \bibinfo{journal}{Proceedings of the National Academy of Sciences}
  \textbf{\bibinfo{volume}{105}}, \bibinfo{pages}{1118} (\bibinfo{year}{2008}).

\bibitem[{\citenamefont{Ronhovde and Nussinov}(2009)}]{24}
\bibinfo{author}{\bibfnamefont{P.}~\bibnamefont{Ronhovde}} \bibnamefont{and}
  \bibinfo{author}{\bibfnamefont{Z.}~\bibnamefont{Nussinov}},
  \bibinfo{journal}{Physical Review E} \textbf{\bibinfo{volume}{80}},
  \bibinfo{pages}{016109} (\bibinfo{year}{2009}).

\bibitem[{\citenamefont{Blondel et~al.}(2008)\citenamefont{Blondel, Guillaume,
  Lambiotte, and Lefebvre}}]{25}
\bibinfo{author}{\bibfnamefont{V.}~\bibnamefont{Blondel}},
  \bibinfo{author}{\bibfnamefont{J.}~\bibnamefont{Guillaume}},
  \bibinfo{author}{\bibfnamefont{R.}~\bibnamefont{Lambiotte}},
  \bibnamefont{and} \bibinfo{author}{\bibfnamefont{E.}~\bibnamefont{Lefebvre}},
  \bibinfo{journal}{Journal of Statistical Mechanics: Theory and Experiment}
  \textbf{\bibinfo{volume}{P10008}} (\bibinfo{year}{2008}).

\bibitem[{\citenamefont{Duch and Arenas}(2005)}]{26}
\bibinfo{author}{\bibfnamefont{J.}~\bibnamefont{Duch}} \bibnamefont{and}
  \bibinfo{author}{\bibfnamefont{A.}~\bibnamefont{Arenas}},
  \bibinfo{journal}{Physical review E} \textbf{\bibinfo{volume}{72}},
  \bibinfo{pages}{027104} (\bibinfo{year}{2005}).

\bibitem[{\citenamefont{Pielou}(1966)}]{17}
\bibinfo{author}{\bibfnamefont{E.}~\bibnamefont{Pielou}},
  \bibinfo{journal}{Journal of theoretical biology}
  \textbf{\bibinfo{volume}{13}}, \bibinfo{pages}{131} (\bibinfo{year}{1966}).

\bibitem[{\citenamefont{Pu et~al.}(2009)\citenamefont{Pu, Wong, Turner, Cho,
  and Wodak}}]{18}
\bibinfo{author}{\bibfnamefont{S.}~\bibnamefont{Pu}},
  \bibinfo{author}{\bibfnamefont{J.}~\bibnamefont{Wong}},
  \bibinfo{author}{\bibfnamefont{B.}~\bibnamefont{Turner}},
  \bibinfo{author}{\bibfnamefont{E.}~\bibnamefont{Cho}}, \bibnamefont{and}
  \bibinfo{author}{\bibfnamefont{S.}~\bibnamefont{Wodak}},
  \bibinfo{journal}{Nucleic acids research} \textbf{\bibinfo{volume}{37}},
  \bibinfo{pages}{825} (\bibinfo{year}{2009}).

\bibitem[{\citenamefont{Zachary}(1977)}]{19}
\bibinfo{author}{\bibfnamefont{W.}~\bibnamefont{Zachary}},
  \bibinfo{journal}{Journal of anthropological research} pp.
  \bibinfo{pages}{452--473} (\bibinfo{year}{1977}).

\bibitem[{\citenamefont{Freeman}(1993)}]{20}
\bibinfo{author}{\bibfnamefont{L.}~\bibnamefont{Freeman}},
  \bibinfo{journal}{Journal of Mathematical Sociology}
  \textbf{\bibinfo{volume}{17}}, \bibinfo{pages}{227} (\bibinfo{year}{1993}).

\bibitem[{\citenamefont{Lucas et~al.}(2006)\citenamefont{Lucas, Arnau, and
  Mar{\'\i}n}}]{21}
\bibinfo{author}{\bibfnamefont{J.}~\bibnamefont{Lucas}},
  \bibinfo{author}{\bibfnamefont{V.}~\bibnamefont{Arnau}}, \bibnamefont{and}
  \bibinfo{author}{\bibfnamefont{I.}~\bibnamefont{Mar{\'\i}n}},
  \bibinfo{journal}{Journal of molecular biology}
  \textbf{\bibinfo{volume}{357}}, \bibinfo{pages}{9} (\bibinfo{year}{2006}).

\bibitem[{\citenamefont{Marco and Mar{\'\i}n}(2009)}]{22}
\bibinfo{author}{\bibfnamefont{A.}~\bibnamefont{Marco}} \bibnamefont{and}
  \bibinfo{author}{\bibfnamefont{I.}~\bibnamefont{Mar{\'\i}n}},
  \bibinfo{journal}{BMC systems biology} \textbf{\bibinfo{volume}{3}},
  \bibinfo{pages}{69} (\bibinfo{year}{2009}).

\bibitem[{\citenamefont{MacArthur}(1957)}]{27}
\bibinfo{author}{\bibfnamefont{R.}~\bibnamefont{MacArthur}},
  \bibinfo{journal}{Proceedings of the National Academy of Sciences of the
  United States of America} \textbf{\bibinfo{volume}{43}}, \bibinfo{pages}{293}
  (\bibinfo{year}{1957}).

\bibitem[{\citenamefont{Newman}(2006)}]{28}
\bibinfo{author}{\bibfnamefont{M.}~\bibnamefont{Newman}},
  \bibinfo{journal}{Proceedings of the National Academy of Sciences}
  \textbf{\bibinfo{volume}{103}}, \bibinfo{pages}{8577} (\bibinfo{year}{2006}).

\end{thebibliography}

\end{document}